\documentclass[aoas,MSNbibl,nameyear,seceqn,dvips]{arximspdf}
\usepackage{dcolumn}
\usepackage{graphicx}

\doi{10.1214/11-AOAS472}
\volume{5}
\issue{3}
\pubyear{2011}
\firstpage{2169}
\lastpage{2196}

\makeatletter
\setattribute{copyright}{owner}{\textup{In the Public Domain}}
\renewcommand{\epsilon}{\varepsilon}
\newcolumntype{d}[1]{D{.}{.}{#1}}
\newcolumntype{e}[1]{D{.}{}{#1}}

\newtheorem{theo}{Theorem}[section]
\newproclaim{defn}{Definition}[section]
\newtheorem{lem}{Lemma}[section]
\newproclaim{rem}[theo]{Remark}

\newcommand{\fraca}[2]{{#1}/{#2}}
\makeatother

\begin{document}
\begin{frontmatter}

\title{Local kernel canonical correlation analysis with application to
virtual drug screening}
\runtitle{Local kernel CCA}

\begin{aug}
\author[A]{\fnms{Daniel} \snm{Samarov}\corref{}\ead[label=e1]{daniel.samarov@nist.gov}},
\author[B]{\fnms{J. S.} \snm{Marron}\ead[label=e2]{marron@email.unc.edu}},
\author[B]{\fnms{Yufeng} \snm{Liu}\thanksref{t1}\ead[label=e3]{yfliu@email.unc.edu}},
\author[C]{\fnms{Christopher} \snm{Grulke}\thanksref{t2}\ead[label=e4]{grulke@email.unc.edu}}
\and
\author[C]{\fnms{Alexander} \snm{Tropsha}\thanksref{t2}\ead[label=e5]{tropsha@unc.edu}}
\runauthor{D. Samarov et al.}
\affiliation{National Institute of Standards and Technology,
University of North Carolina, University of North Carolina,
University of North Carolina and University~of~North Carolina}
\address[A]{D. Samarov\\
Statistical Engineering Division \\
Information Technology Laboratory \\
National Institute of Standards \\
\quad and Technology\\
Gaithersburg, Maryland 20878\\
USA\\
\printead{e1}} 
\address[B]{J. S. Marron\\
Y. Liu\\
Department of Statistics and\\
 \quad Operations Research\\
University of North Carolina\\
Chapel Hill, North Carolina 27599-3260\\
USA\\
\printead{e2}\\
\phantom{\textsc{E-mail:}\ }\printead*{e3}}
\address[C]{C. Grulke\\
A. Tropsha\\
School of Pharmacy\\
University of North Carolina\\
Chapel Hill, North Carolina 27599-3260\\
USA\\
\printead{e4}\\
\phantom{\textsc{E-mail:}\ }\printead*{e5}}
\end{aug}
\thankstext{t1}{Supported in part by NSF Grant DMS-07-47575 and NIH
Grant NIH/NCI R01 CA-149569.}
\thankstext{t2}{Supported in part by NIH Grant GM066940.}

\received{\smonth{8} \syear{2009}}
\revised{\smonth{3} \syear{2011}}

%
\begin{abstract}
Drug discovery is the process of identifying compounds which have potentially
meaningful biological activity. A major challenge that arises is that the
number of compounds to search over can be quite large, sometimes
numbering in
the millions, making experimental testing intractable. For this reason
computational methods are employed to filter out those compounds which
do not
exhibit strong biological activity. This filtering step, also called virtual
screening reduces the search space, allowing for the remaining
compounds to be
experimentally tested.

In this paper we propose several novel approaches to the problem of virtual
screening based on Canonical Correlation Analysis (CCA) and on a kernel-based
extension. Spectral learning ideas motivate our proposed new method called
Indefinite Kernel CCA (IKCCA). We show the strong performance of this approach
both for a toy problem as well as using real world data with dramatic
improvements in predictive accuracy of virtual screening over an existing
methodology.
\end{abstract}

%

\end{frontmatter}

\section{Introduction}
\label{SECINTRODUCTION}
%
Computer-Aided Drug Discovery (CADD) is an area of research that is concerned
with the identification of chemical compounds that are likely to
possess specific
biological activity, that is, the ability to bind certain target
biomolecules such as
proteins. CADD approaches are employed in order to prioritize molecules in
commercially available chemical libraries for experimental biological screening.
The prioritization of molecules is critical since these libraries frequently
contain many millions of molecules making experimental testing
intractable. The
process of using computational methods to filter out those compounds
which are
not expected to exhibit strong biological activity is called virtual screening.

Computational methods have been used extensively to assist in
experimental drug
discovery studies. In general, there are two major computational drug discovery
approaches, ligand based and structure based. The former is used when the
three-dimensional structure of the drug target is unknown but the information
about a reasonably large number of organic molecules active against a specific
set of targets is available. In this case, the available data can be studied
using cheminfomatic approaches such as Quantitative Structure-Activity
Relationship (QSAR) modeling [for a review of QSAR methods see A.
Tropsha, in
\citet{abraham2003}]. In contrast, the structure-based methods rely on the
knowledge of three-dimensional structure of the target protein,
especially its
active site; this data can be obtained from experimental structure elucidation
methods such as X-ray or Nuclear Magnetic Resonance (NMR) or from
modeling of
protein three-dimensional structure.

Virtual screening is one of the most popular structure-based CADD approaches
where, typically, three-dimensional protein structures are used to
discover small
molecules that fit into the active site (a process referred to as
docking) and
have high predicted binding affinity (scoring). Traditional docking
protocols and
scoring functions rely on explicitly defined three-dimensional
coordinates and
standard definitions of atom types of both receptors and ligands. Albeit
reasonably accurate in some cases, structure-based virtual screening approaches
are for the most part computationally inefficient [\citet{warren2006}].
As a
result of computational inefficiency there is a limit to the number of
compounds which can reasonably be screened by these methods. Furthermore,
recent extensive studies into the comparative accuracy of multiple available
scoring functions suggest that accurate prediction of binding
orientations and
affinities of receptor--ligand pairs remains a formidable challenge
[\citet{kitchen2004}]. Yet millions of compounds in available chemical
databases and billions of compounds in synthetically feasible chemical
libraries are available for virtual screening calling for the
development of
approaches that are both fast and accurate in their ability to identify a
small number of viable and experimentally testable computational hits.

Recently, we introduced a novel structure-based cheminformatic workflow
to search
for Complimentary Ligands Based on Receptor Information
(CoLiBRI) [\citet{colibri2006}]. This novel computational drug
discovery strategy
combines the strengths of both structure-based and ligand-based
approaches while
attempting to surpass their individual shortcomings. In this approach, we
extract the structure of the binding pocket from the protein and then represent
both the receptor active site and its corresponding ligand in the same
universal,
multidimensional chemical descriptor space (note that in principle, the
descriptors used for receptors and ligands do not have to be the same,
and we
will be exploring the use of different descriptor types in future
studies). We
reasoned that mapping of both binding pockets and corresponding ligands
onto the
same multidimensional chemistry space would preserve the complementary
relationships between binding sites and their respective ligands. Thus,
we expect
that ligands binding to similar active sites are also similar. In
cheminformatics
applications, the similarity is described quantitatively using one of the
conventional metrics, such as Manhattan or Euclidean distance in
multidimensional
descriptor space. Thus, the chief hypothesis in CoLiBRI is that the relative
location of a novel binding site with respect to other binding sites in
multidimensional chemistry space could be used to predict the location
of the
ligand(s) complementary to this site in the ligand chemistry space. After
generation of descriptors, the dataset is split into training and test
sets and
then variable selection is carried out to generate models optimizing this
complementarity between the binding pocket and ligand spaces. These
models are
then applied to a binding pocket in a~protein of interest to generate a
predicted
virtual ligand point which is used as a query in chemical similarity
searches to
identify putative ligands of the receptor in available chemical databases.
In this paper, we build upon the work of \citet{colibri2006} to
develop a substantially more advanced and efficient version of CoLiBRI.

The problem can be generally stated as follows: for a set of $n$ known
protein--ligand pairs, with $d_X$ and $d_Y$ descriptors, respectively, given a
new protein
we want to be able to predict what ligand(s) will bind to it. Two
virtual drug
screens will be used as a benchmark for testing the methods discussed and
developed here:
\begin{longlist}[(2)]
\item[(1)] A set of 800 chemically and functionally diverse protein--ligand pairs
obtained from the Protein Data Bank (PDB) database on experimentally
measured binding affinity (PDBbind) [\citet{pdbbind2004}]. These
compounds are
described by a set of 150 chemical descriptors. These descriptors include
information related to the electronic attributes, hydrophobicity and steric
properties of the compounds. For a more detailed discussion on the different
types of chemical descriptors, see \citeauthor{cheminfo2009A}
(\citeyear{cheminfo2009A,cheminfo2009B}).
We will refer to this data set as the 800
Receptor--Ligand pairs (RLP800) data. Results and further details on
this and
two additional data sets can be found in Section \ref{SECResults}.
\item[(2)] The World Drug Index (WDI) [\citet{wdi2004}] database which contains
approximately 54,000 drug candidates (ligands). Each compound in the
WDI is
described by the same set of 150 chemical descriptors as the RLP800 data.
\end{longlist}

The accuracy of our prediction is based on how close, in Euclidean distance,
our prediction is to the actual ligand. This is then compared against the
distances of all of the ligands in the space to the actual ligand.\vadjust{\eject} A standard
measure of predictive accuracy used in the QSAR literature
[\citet{tropsha2003}, \citet{colibri2006}] is based on ranking these distances,
from smallest to largest. Defining $r_i$ to be the rank of our
prediction of
test ligand $i$, model performance is defined as the average rank over
each of
the new points we are trying to predict, $\bar{r} = \sum_{i=1}^{n_T} r_i$.
This criterion reflects the average size of the search space needed to find
each compound. Here $n_T$ denotes the number of new (i.e., test)
ligands we are
predicting.

\begin{figure}

\includegraphics{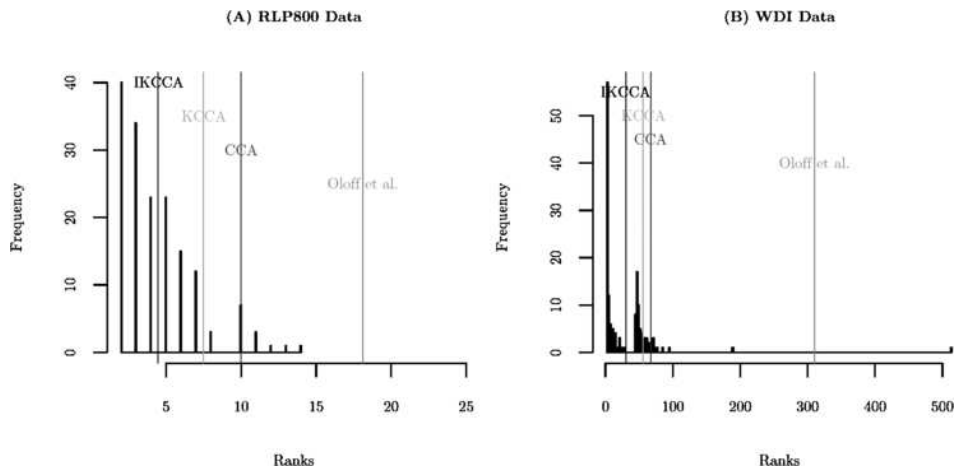}

\caption{\textup{(A)}  A histogram showing the IKCCA ranks, $r_i$,
resulting from prediction on the test data from the RLP800 dataset.
\textup{(B)}   Performance on the WDI data.}
\label{FIGIKCCA}
\end{figure}

The $\bar{r}$ effectiveness of the methods studied and developed here is
illustrated in Figure \ref{FIGIKCCA}. Figure \ref{FIGIKCCA}(A) is a
histogram of the ranks, $r_i$ for our novel method which is a variant of
Canonical Correlation Analysis (CCA) we call \textit{Indefinite Kernel CCA}
(IKCCA) (Section \ref{SECIKCCA}), on the RLP800 data.

The previous state of the art for these data sets is $\bar{r}_{\mathrm{OLOFF}}$ (the
vertical line furthest to the right labelled Oloff et al.) which are
larger by a
factor of 5 to 10 as compared to CCA (Section \ref{SECCCA}) and
its improvements, KCCA (Section \ref{SECKCCA}) and IKCCA.

As we were primarily interested in comparing our results against those of
\citet{colibri2006} we did not look into other performance metrics
other than
mean rank. However, it would be interesting to pursue other,
potentially more
relevant measures of binding affinity such as Kd, Ki and
IC50 as was done by \citet{witten2009B}, where CCA is linked to these
performance measures.

While not discussed in this paper, an important unresolved issue in this
cheminformatic-based approach to the prediction of protein--ligand
binding is the
selection of meaningful chemical descriptors. The type of chemical
descriptors used can have a drastic effect on the predictive accuracy of
an algorithm. One possible approach\vadjust{\eject} to addressing this issue would be
to use a
recently developed method called Sparse CCA (SCCA),
\citet{hardoon2007}, \citet{park2009}, \citet{witten2009A} and \citet
{witten2009B}.
SCCA uses a lasso-like approach to identify sparse linear combinations
of two
sets of variables that are highly correlated with each other. An
approach based
on SCCA to the prediction of protein--ligand binding may prove to be quite
useful in resolving some of the issues arising from chemical descriptor
selection.

In Section
\ref{SECResults} we present results and details on the RLP800 data
set as well
as on two additional data sets.
In Sections \ref{SECCCA} and \ref{SECKCCA} we outline CCA and KCCA,
respectively. In Section \ref{SECIKCCA} we propose a new method,
IKCCA, which
encompasses nonpositive semi-definite (PSD) kernels (i.e., indefinite kernels),
specifically we consider a class of kernels related to the Normalized Graph
Laplacian used in Spectral Clustering. Finally, in Section \ref{SECPREDICTION}
we show how prediction of a new ligand is done using CCA (and its variants).

%
\subsection{Additional drug discovery results}
\label{SECResults}
%
In addition to the real data results discussed in Section \ref
{SECINTRODUCTION},
we also tested our method on two additional data sets [which we refer
to as
Experimental Settings (ES), the reason for which will become clearer in what
follows]. These data (including the RLP800 data%
)
are subsets of a collection of 1\mbox{,}300 complexes taken from
PDBBind [\citet{pdbbind2004}]. These 1\mbox{,}300 complexes are referred to as
the \textit{Refined Set} (RS), a set of entries that meet a defined set of criteria
regarding
crystal structure quality. A representative subsample of 195 of the
complexes is
called the \textit{Core Set} (CS). This is a collection of complexes
selected by
clustering the RS into 65 groups using protein sequence similarity and retaining
only 3 complexes from each cluster.

The three experimental settings considered are denoted by ES I [this
experimental setting was used in \citet{colibri2006}], ES II and ES
III. In
each of these experimental settings the RS and CS complexes are
separated into
training and testing sets in such a way as to test different aspects of our
model. In ES I the 637 training and 163 test complexes are randomly sampled
from the RS. ES I is meant to provide a general test of our models
performance. In ES II the training (153 complexes) and testing (36 complexes)
sets are sampled from the CS in such a way that the various protein families
in the CS are well represented in both. This separation is meant to
test the
performance of our CCA-based methods when the sample size is small. Finally,
in ES III the testing set (162 complexes) is composed of proteins which are
under represented in the training set (1\mbox{,}006 complexes). This is meant
to test
our methods ability to correctly identify novel complexes.

\begin{table}
\tabcolsep=0pt
\caption{This table summarizes the performance of the method
discussed in
Oloff et~al. (\protect\citeyear{colibri2006}) as well as the methods developed
in this paper on the
CoLiBRI, Name and Cluster data sets. The columns labeled ``Train'' and
``Test'' correspond to the number of training and testing samples for a given
data set. The column labeled ``Embed'' corresponds to the total number of
ligands against which our prediction is to be ranked. The remaining columns
correspond to the method used and the average rank performance of that method.
Note that as the method used in Oloff et~al. (\protect\citeyear{colibri2006})
failed to provide useful
results for the Name and Cluster data sets, no results were reported.
In all
cases IKCCA, using the NGL kernel, outperformed the other methods. In all
cases the CCA-based methods provide considerable improvement over the previous
approach}
\label{TABLERESULTS}
\begin{tabular*}{\textwidth}{@{\extracolsep{\fill}}lcce{3.0}cd{3.1}d{3.1}d{2.2}d{2.1}@{}}
\hline
\textbf{Setting} & & \multicolumn{1}{c}{\textbf{Train}} & \multicolumn{1}{c}{\textbf{Test}} &
\multicolumn{1}{c}{\textbf{Embed}} & \multicolumn{1}{c}{\textbf{Oloff}} &
\multicolumn{1}{c}{\textbf{CCA}}
& \multicolumn{1}{c}{\textbf{KCCA}} &
\multicolumn{1}{c@{}}{\textbf{IKCCA}} \\
\hline
ES I & RS & \hphantom{1,}637 & 163 & \hphantom{54\mbox{,}}800 &
18.1 & 10 & 7.5 & 4.5 \\
& RS${}+{}$WDI & & & 54\mbox{,}121 & 310 & 67 & 56 & 30 \\
[3pt]
ES II & RS & \hphantom{1\mbox{,}}153 & 36 & \hphantom{54\mbox{,}}189 & \multicolumn{1}{c}{NA}
& 8 & 13.75 & 3.5 \\
& RS${}+{}$WDI & & & 53\mbox{,}994 & \multicolumn{1}{c}{NA} & 275.1 & \multicolumn{1}{c}{1\mbox{,}558}
& 92.9 \\
[3pt]
ES III & RS & 1\mbox{,}006 & 162 & \hphantom{5}1\mbox{,}168& \multicolumn{1}{c}{NA} & 11.9
& 7.4 & 4.4 \\
& RS${}+{}$WDI & & & 54\mbox{,}120 & \multicolumn{1}{c}{NA} & 53 & 24.3 & 18.2 \\
\hline
\end{tabular*}
\vspace*{5pt}
\end{table}

A note on how we use the training and testing sets: the tuning
parameters for
our model are selected, as discussed in Section \ref{SUBSECTUNEPARAMS},
using only the training set. Once tuning parameters have been selected,
prediction on the testing set is then performed. This is meant to test the
models performance on as-yet unobserved complexes.\vadjust{\eject}

The results for each of these experimental settings is summarized in Table~\ref{TABLERESULTS}. The columns labeled ``Train'' and ``Test''
correspond to
the size of the training/testing sets for each particular experimental
setting. The column labeled ``Embed'' corresponds to the total number of
ligands against which our prediction is to be ranked. The remaining columns
correspond to the method used and the average rank performance (defined in
Section \ref{SECINTRODUCTION}) of that method.
The second row, second column in each cell labeled ``RS${}+{}$WDI'' shows the
results for each method on the Reduced Set plus the World Drug
Index. These results are meant to more accurately mimic an actual drug screen
by having a larger test set to search against. As the method used in
\citet{colibri2006} failed to provide useful results for the ES II and
ES III
experimental settings, no results are reported here. Generally
speaking, in
all cases IKCCA, using the NGL kernel, outperformed the other methods.
All the
CCA-based methods provide considerable improvement over the previous approach.

Looking a bit closer at the results it is interesting to note that
while all
three CCA-based methods performed worse on the ES II data, KCCA had the
largest drop in performance. This can be seen by comparing the average rank
performance against the total number of ligands we are searching
against. The
decrease in performance in all cases more than likely has to do with
the small
size of the training set. In the case of KCCA, its considerable
decrease in
performance, we suspect, may have to do with not having a large enough
training sample to reliably select the bandwidth parameter $\sigma$.
For IKCCA
the adaptive nature of the local kernel is probably what allows it to perform
well in the low sample size setting.

\section{Canonical correlation analysis}
\label{SECCCA}
CCA [\citet{Hotelling1936}] naturally lends itself to the problem of predicting
the binding between proteins and ligands. This can be understood for the
following reasons: first, traditional methods of prediction, for
example, regression,
assume a direction of dependence between the variables to be predicted
and the
predictive variables. Here we have a symmetric, not causal, type of
relationship: the binding between a~protein and its ligand is inherently
co-dependent. Second, in addition to capturing the dependence structure
we are
looking to model, CCA is well suited to the type of prediction we are
interested in performing. To understand this, consider the following
(see also
Section~\ref{SECTOYEX1} for a more detailed discussion). The
objective of
CCA is to find directions in one space, and directions in a second
space such
that the correlation between the projections of these spaces onto their
respective directions is maximized. These directions are commonly
referred to
as canonical vectors. Let us assume that a~set of directions are found
so that
the corresponding projections of proteins and of ligands are strongly
correlated. Predicting a new ligand given a new protein would begin with
projecting the new protein into canonical correlation space. Then, assuming
the same correlation structure holds for this new point, prediction of
the new
ligand would amount to interpolating its location in ligand space based
on the
location of the protein in protein space. This will be discussed in greater
detail in Section \ref{SECPREDICTION}. Next we provide a brief
discussion on
the details of CCA and KCCA.

\subsection{Canonical correlations}
\label{SUBSECCCA}
Let $\mathbf{x}_i\in\mathbb{R}^{d_X}$ and $\mathbf{y}_i\in\mathbb
{R}^{d_Y}$,
$i=1,\ldots,n,$ denote a protein--ligand pair. The sample of pairs is collected
in matrices $\mathbf{X}\in\mathbb{R}^{n\times d_X}$ and
$\mathbf{Y}\in\mathbb{R}^{n\times d_Y}$ with $\mathbf{x}_i$ and
$\mathbf{y}_i$
as the descriptors for a row.

The objective of CCA is to find the linear combinations of the columns of
$\mathbf{X}$ (proteins), say $\mathbf{X}\mathbf{w}_X$ and the linear
combinations of the columns of~$\mathbf{Y}$ (ligands), say
$\mathbf{Y}\mathbf{w}_Y$ such that the correlation,
$\operatorname{corr}(\mathbf{X}\mathbf{w}_X,\mathbf{Y}\mathbf{w}_Y)$ is maximized.
Without loss of generality assume that the matrices $\mathbf{X}$ and
$\mathbf{Y}$ have been mean centered. Letting $\mathbf{S}_{XX} =
\mathbf{X}^T\mathbf{X}$, $\mathbf{S}_{YY} = \mathbf{Y}^T\mathbf
{Y}$ and
$\mathbf{S}_{XY} = \mathbf{X}^T\mathbf{Y}$ the CCA optimization
problem is
%
\begin{eqnarray}\label{EQCCA}
&&\rho =
\max_{\mathbf{w}_X,\mathbf{w}_Y}\operatorname{corr}
(\mathbf{X}\mathbf{w}_X,\mathbf{Y}\mathbf{w}_Y)    \nonumber
\\
&&\mbox{subject to,}\\
&&\mathbf{w}_X^T\mathbf{S}_{XX}\mathbf{w}_X =
\mathbf{w}_Y^T\mathbf{S}_{YY}\mathbf{w}_Y = 1.\nonumber
\end{eqnarray}
 Subsequent directions are found by imposing the additional
constraints\vspace*{-1pt}\break $(\mathbf{w}_X^i)^T\times \mathbf{S}_{XX}\mathbf{w}_X^j =
(\mathbf{w}_Y^i)^T\mathbf{S}_{YY}\mathbf{w}_Y^j =
(\mathbf{w}_X^i)^T\mathbf{S}_{XY}\mathbf{w}_Y^j = 0$ for $i\neq j$
and\break
$(\mathbf{w}_X^i)^T\mathbf{S}_{XX}\mathbf{w}_X^i =
(\mathbf{w}_Y^i)^T\mathbf{S}_{YY}\mathbf{w}_Y^i = 1$, $i, j =
1,\ldots,p$, $p = \operatorname{min}(d_X,d_Y)$.

In order to avoid issues arising from multicollinearity and singularity
of the
covariance matrices we impose a penalty [\citet{vinod1976}] on the directions~$\mathbf{w}_X$ and $\mathbf{w}_Y$ so that the constraints in (\ref{EQCCA})
are modified to be
%
\begin{equation}\label{EQRCCA}
\mathbf{w}_X^T\mathbf{S}_{XX}\mathbf{w}_X + \kappa\mathbf
{w}_X^T\mathbf{w}_X =
\mathbf{w}_Y^T\mathbf{S}_{YY}\mathbf{w}_Y + \kappa\mathbf
{w}_Y^T\mathbf{w}_Y =
1,
\end{equation}
 where $\kappa\in\mathbb{R}$ is a regularization parameter.

The predictive accuracy of this approach was discussed in Section
\ref{SECINTRODUCTION}, with results summarized in Figure \ref{FIGIKCCA}.
Recall that the  lines in these figures labeled CCA correspond to the average
predicted rank using CCA, which improved upon \citet{colibri2006}
shown by the lines labeled a such.

\subsubsection{The geometry of CCA}
\label{SUBSECGEOMCCA}
An appealing aspect of CCA is its intuitive geometric interpretation
[\citet{Anderson2003} and \citet{GeomKCCA}]. A~geometric perspective
lends itself
to a better understanding of the general behavior of CCA, and provides
further evidence of
its applicability to the protein--ligand matching problem.

Taking a closer look at the $i$th canonical correlation, $\rho_i$, $i =
1,\ldots,p$ ($p=\operatorname{min}(d_X,d_Y)$), in the optimization problem
shown in
(\ref{EQCCA}), it can be seen that this quantity is in fact equal to the
cosine of the angle between $\mathbf{a}_X^i = \mathbf{X}\mathbf
{w}^i_X$ and
$\mathbf{a}_Y^i = \mathbf{Y}\mathbf{w}^i_Y$ ($\mathbf{a}_X^i$ and
$\mathbf{a}_Y^i$ are commonly referred to as canonical variates). With
this in
mind maximizing the cosine (i.e., correlation) can equivalently be
thought of
as minimizing the angle between $\mathbf{a}_X^i$ and $\mathbf{a}_Y^i$.
Furthermore, it can be shown that minimizing the angle is equivalent to
minimizing the distance between pairs of canonical variates,
\[
\min_{\mathbf{w}^i_X,\mathbf{w}^i_Y}
\|\mathbf{X}\mathbf{w}^i_X - \mathbf{Y}\mathbf{w}^i_Y\|^2
\]
 subject to the constraints described in (\ref{EQCCA}). Note that
viewed in this way, in canonical correlation space, this amounts to
finding a
system of coordinates such that the distance between coordinates is minimized.
This is a sense in which CCA is an appropriate approach to the protein--ligand
matching problem.

As will be seen in Sections \ref{SECKCCA} and \ref{SECIKCCA}, this geometric
interpretation of CCA extends naturally to KCCA and IKCCA. Note that the
regularized variant of CCA does not have the same geometric interpretation,
nonetheless viewing regularized CCA in this manner still provides useful
insight into its behavior.

\subsection{Toy examples}
\label{SECTOYEX}

\subsubsection{Toy example 1: Motivating CCA}
\label{SECTOYEX1}
\begin{figure}

\includegraphics{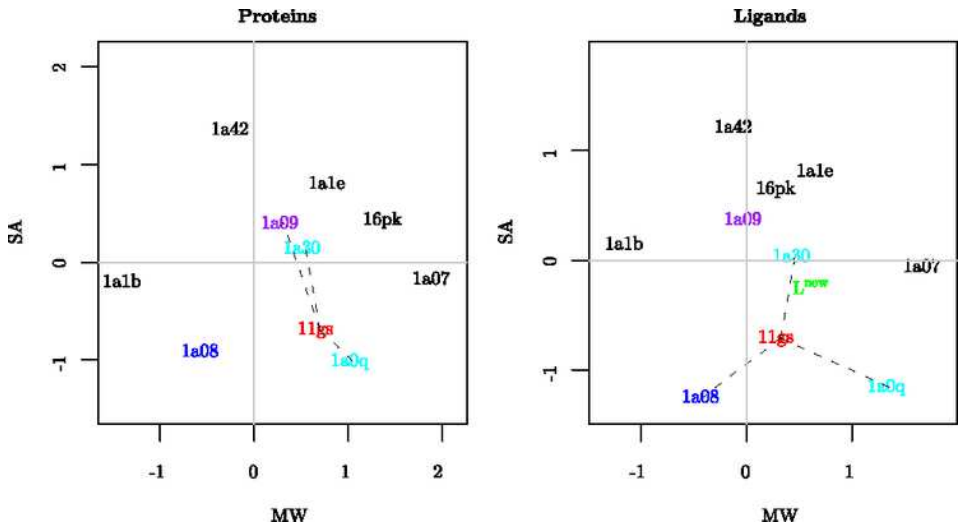}

\caption{Toy example data. The points highlighted in red correspond to the
protein--ligand pair 11gs, and the points connected to it by dashed
black lines
are its three nearest neighbors in each space. The observations
highlighted in
cyan are neighbors in both spaces, and those highlighted in blue and purple
are neighbors only in the protein, and ligand spaces, respectively. The green
point ${L}^{\mathrm{new}}$ in the ligand space corresponds to a weighted average
(discussed in Section \protect\ref{SECPREDICTION}) of the cyan points and the purple
point, that is, of the nearest neighbors of 11gs in the protein space.}
\label{plotex1}
\end{figure}

Consider the protein--ligand\break \mbox{matching} problem as outlined above. For
this toy
example we set $n=10$ and $d=2$. Suppose the descriptors for this toy example
are Molecular Weight (MW) and Surface Area (SA) of the molecule. Recall that
each row of~$\mathbf{X}_{(10\times2)}$ and each row of $\mathbf
{Y}_{(10\times
2)}$ corresponds to an observation, a~protein or a ligand, respectively, and
the columns correspond to the descriptors MW and SA. The pairs are identified
by a unique label, corresponding to IDs from the Protein Data Bank (PDB)
(\href{http://www.pdb.org}{www.pdb.org}). Figure \ref{plotex1} shows the two toy data sets.

From Figure \ref{plotex1} it can be seen that the distribution of
points in
the two spaces are quite similar in the sense that the location of
corresponding points in the two spaces are close. The points connected
to 11gs
(red) by dashed black lines are its three nearest neighbors. The cyan points
are neighbors shared in both spaces and the blue and purple points are
mismatched. Two of three neighbors are shared in common (in the Euclidean
sense).

Consider the case where the red point in ligand space is not observed
and the
task is to predict its value. Using the weighted average (see Section
\ref{SECPREDICTION} for details on the derivation of the weights) of
the points
in ligand space that correspond to the nearest neighbors of the point
11gs in the
protein space (points highlighted in cyan and purple in ligand space)
would yield
a relatively poor prediction despite the strong apparent similarity
between the
two distributions of points.

\begin{figure}

\includegraphics{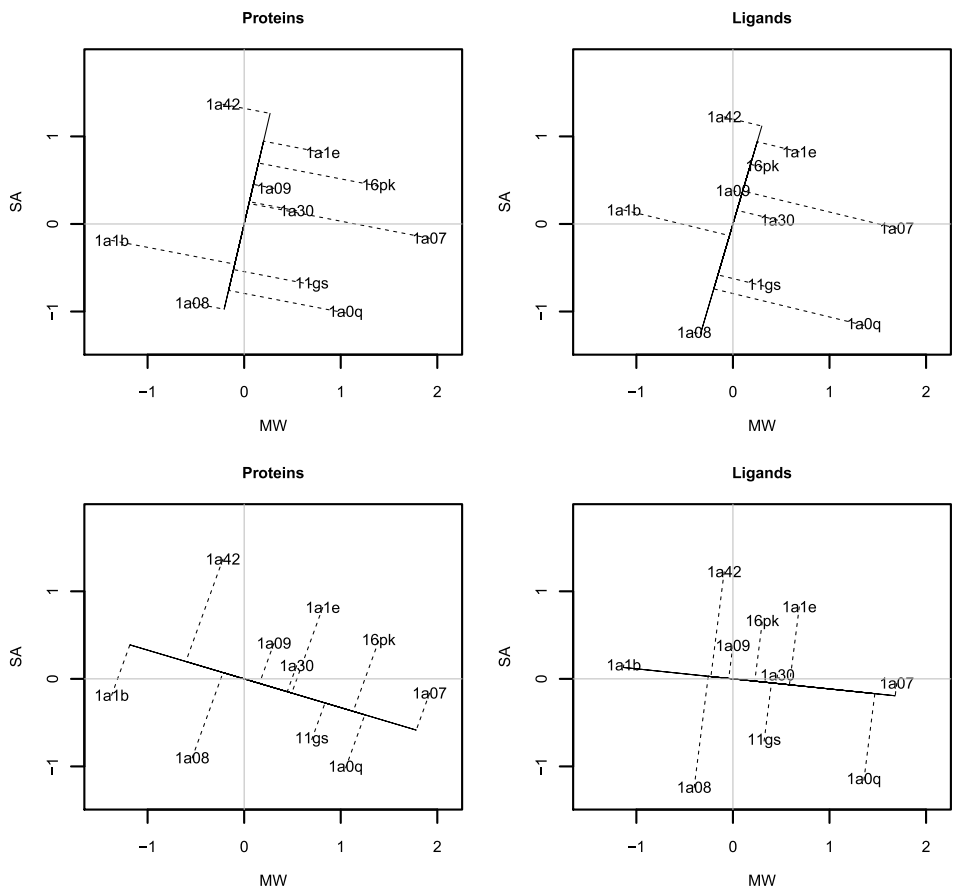}

\caption{The direction vectors and the projected value of each point. The
top row of plots shows the first direction vector
and the projections
onto it. The bottom row of plots show the second direction vector
and the projection onto it.}\label{ccaprojdir}
\vspace*{12pt}
\end{figure}

Next suppose that instead of carrying out the prediction of a new
ligand in
the original data space we carry out our prediction in canonical correlation
space. Solving for $\mathbf{w}_X$ and $\mathbf{w}_Y$ in (\ref
{EQCCA}), gives
us the canonical vectors shown in Figure \ref{ccaprojdir}. What is important
to notice is how the distribution of points along the first  and second
 canonical directions in both protein and ligand space are quite
similar. This is due to the property of alignment that arises naturally from
maximizing the correlation.

\begin{figure}

\includegraphics{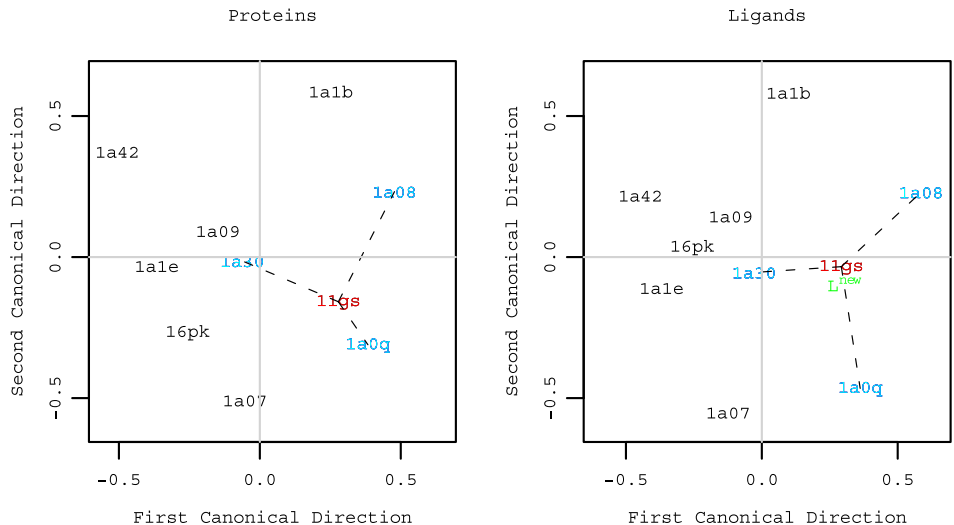}

\caption{Projection of the data in Figure \protect\ref{plotex1} onto the first
and second canonical vectors. In contrast to Figure \protect\ref{plotex1},
the point
11gs now shares the same neighbors in both spaces and the predicted
value in
green is much closer to the actual value.}\label{motivateprojplot}
\end{figure}

Figure \ref{motivateprojplot} shows the projections of the data onto the
first two canonical vectors (note that separate directions are found in
protein and ligand space). We can see that with the slight modification in
alignment that has resulted from the CCA projections, the point 11gs now
shares the same neighbors in both spaces. In particular note that now the
predicted value in the projected ligand space is closer to the actual value
(again using the weighted average).

This example was deliberately chosen to illustrate the case where CCA is
effective. However, in most cases the relationship between points in different
spaces may be far more complicated, as we now illustrate.

\subsubsection{Toy example 2: CCA challenge}
\label{SECTOYEX2}
\begin{figure}

\includegraphics{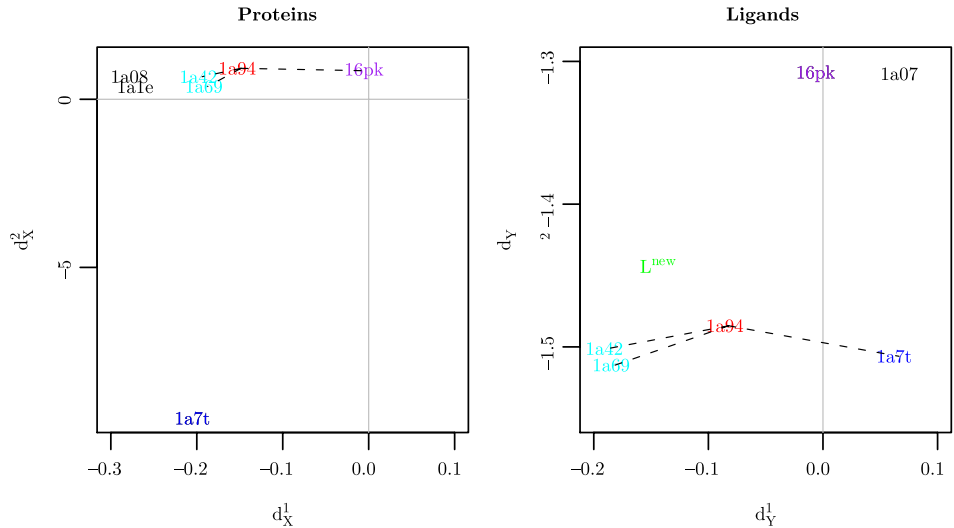}

\caption{A plot of the data generated such that the underlying
relationship between points is nonlinear. The observation highlighted
in red,
1a94, is the new observation which we are trying to predict. The points joined
to it by dashed black lines are its nearest neighbors. The points highlighted
in cyan correspond to points that are nearest neighbors of 1a94 in both
spaces. Points highlighted in purple and blue correspond to points that are
only neighbors in either protein or ligand space, respectively. The point
labeled $L^{\mathrm{new}}$ in ligand space corresponds to a weighted average of the
points 1a08, 1a09 and 1a1b, that is, the nearest neighbors of the point
1a94 in
protein space.}
\label{datasimpleKernel}
\end{figure}

We now consider an example where the relationship between spaces is more
complex. Suppose that we have the same general framework as in Section
\ref{SECTOYEX1} but rather than having both protein and ligand space
characterized by MW and SA, we now have that the space of proteins has
descriptors $d_X^1$ and $d_X^2$ and that the space of ligands has descriptors
$d_Y^1$ and $d_Y^2$, shown in Figure \ref{datasimpleKernel}. As
before the
observation highlighted in red, 1a94, corresponds to a new protein whose
corresponding ligand we are trying to predict. The point highlighted in cyan
is one of the 3-nearest neighbors of 1a94 in both spaces. Those points
highlighted in purple (and blue) are nearest neighbors in only the protein
(and ligand) spaces, respectively. The point $L^{\mathrm{new}}$ in the ligand space,
highlighted in green is a weighted average of the nearest neighbors of the
point 1a94 in protein space. Using $L^{\mathrm{new}}$ as a prediction of the new ligand
would not provide a particularly accurate prediction.

\begin{figure}

\includegraphics{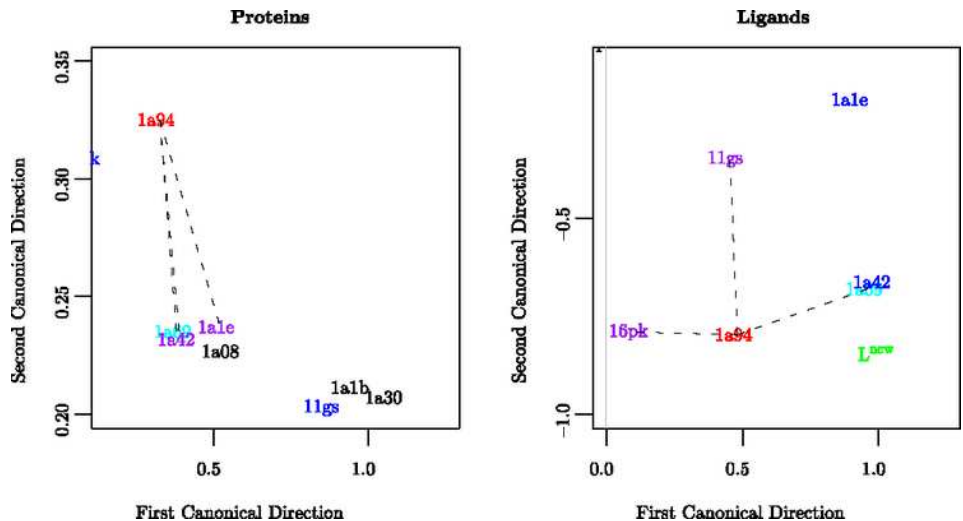}

\caption{A plot of the data projected onto the first two canonical
vectors in both protein and ligand spaces. The directions found by standard
CCA do not provide a good alignment between the two
spaces.}\label{ccsimpleKernel}
\end{figure}

As before, we use CCA to try and find a linear combination of the descriptors
which best align the two spaces. Figure \ref{ccsimpleKernel} is a
plot of the
projections onto the first and second canonical variates in protein and ligand
space. The color scheme is the same as in Figure \ref
{datasimpleKernel}. As
can be seen, standard CCA does not seem to be able to find a good alignment
between the two spaces, which is confirmed by the relatively low values
of the
canonical correlations, 0.79 and 0.54, respectively, for the first and second
directions.

In Section \ref{SECKCCA} we show how mappings into a kernel induced feature
space can be used to improve prediction. This will lead to our
discussion of
KCCA.

\section{Kernel canonical correlation}
\label{SECKCCA}

\subsection{Toy example 2: CCA challenge (motivating KCCA)}
\label{exfeatmap}
\begin{figure}

\includegraphics{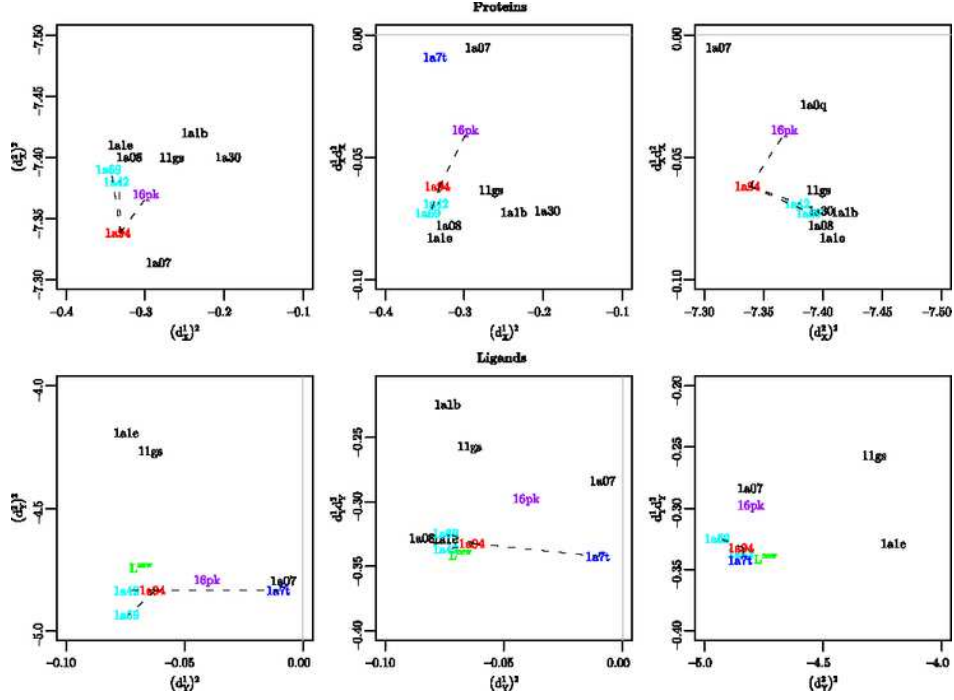}

\caption{A plot of protein and ligand data in feature space. The color
scheme is the same as in Figure \protect\ref{datasimpleKernel}. Looking at
the plots
on the top and bottom (corresponding to protein and ligand space, respectively),
the overall correspondence between points in protein space and ligand
space is
much better than in the original (object) space. This improved mapping will
allow CCA to do a better job aligning the two
spaces.}\label{kernsimpleKernelreceptor}
\end{figure}

Returning to the example in Section \ref{SECTOYEX2}, suppose it is believed
that some type of functional relationship exists between the descriptors
across spaces that is best characterized by looking at the second order
polynomials of the descriptors within each space, that is,
%
\begin{eqnarray}\label{simpleFeatMap}
  \Phi_X\dvtx  (d_X^1, d_X^2) &\rightarrow&((d_X^1)^2,(d_X^2)^2,d_X^1d_X^2),
\nonumber
\\[-8pt]
\\[-8pt]
  \Phi_Y\dvtx  (d_Y^1, d_Y^2) &\rightarrow&((d_Y^1)^2,(d_Y^2)^2,d_Y^1d_Y^2).
\nonumber
\end{eqnarray}
 Figure \ref{kernsimpleKernelreceptor} shows plots of
proteins and
ligands embedded into this three dimensional space. As can be seen
there are
now two neighbors shared in common between spaces (colored in cyan).
Furthermore the prediction of the new observation, $L^{\mathrm{new}}$ (in green)
by a
weighted average of its three nearest neighbors in feature space is, by
comparison, much closer to the actual value than the corresponding prediction
in object space.

\begin{figure}

\includegraphics{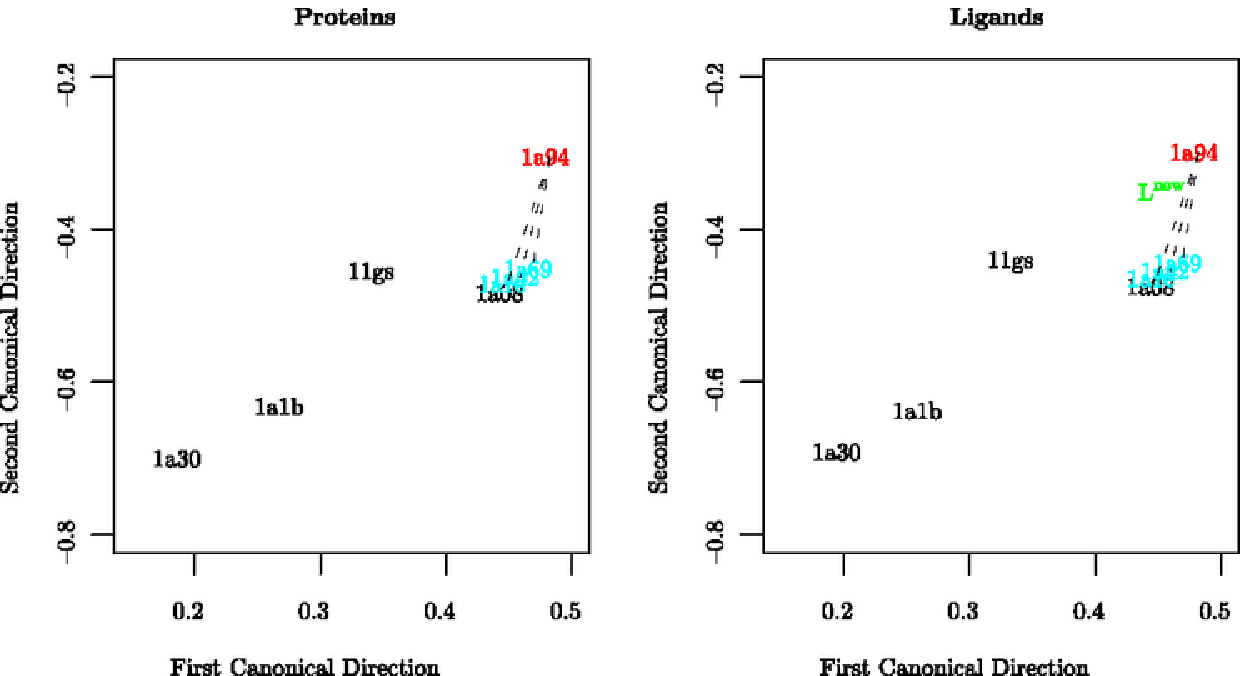}

\caption{This is a plot of the projection of the data in protein and
ligand feature space onto the first and second canonical vectors (note
that we
do not project onto the third canonical vector as the corresponding canonical
correlation is relatively small, approximately 0.2). As can be seen,
not only
does the new observation 1a94 (red) have three neighbors in common in
both protein
and ligand space but the prediction of the new ligand using a weighted
average, $L^{\mathrm{new}}$ highlighted in green on the plot on the right is
close to
the actual value of 1a94.}\label{kernCcsimpleKernelreceptor}
\vspace*{-3pt}
\end{figure}

As before CCA is used on this transformed data, now in feature space,
to align
the space of proteins and ligands. Figure \ref{kernCcsimpleKernelreceptor}
shows a plot of the projected data. Note that now both the new protein
and its
ligand (highlighted in red) share three neighbors and that the
distribution of
points within each of the spaces is quite similar. The quality of the
alignment is further confirmed by looking at the canonical correlation values
which are near 1 for each of the first two directions. Since the value
of the
third canonical correlation is considerably smaller (approximately 0.2) we
only project onto the first two directions.

It is worth noting that, as a result of overfitting, the kernel canonical
correlation values can sometimes be artificially large due to strong
correlation between features in kernel space. Regularization methods for
helping to control these effects in the kernel case will be discussed in
Section \ref{SUBSECKCCA}.

In general, finding explicit mappings such as those in (\ref{simpleFeatMap}) is
impractical or simply not possible as in some cases this would require an
infinite dimensional feature space. As we will see in the following section,
kernels allow us to avoid such issues.

\subsection{Kernel canonical correlation analysis}
\label{SUBSECKCCA}
KCCA [Bach and Jordan\break (\citeyear{KICA2002}), \citet{fyfeLai2000},
\citet{CCAOverview2004}]
extends CCA
by finding directions of maximum correlation in a kernel induced feature
space. Let $\phi_X$ and $\phi_Y$ be the feature space maps for
proteins and
ligands, respectively. The sample of pairs, now mapped into feature
space, are
collected in matrices $\Phi_X$ and $\Phi_Y$ with $\phi_X(\mathbf
{x}_i)$ and
$\phi_Y(\mathbf{y}_i)$ as their respective row elements. The
objective, as
before, is to find linear combinations, $\Phi_X\mathbf{w}_X$ and
$\Phi_Y\mathbf{w}_Y$ such that the correlation,
$\operatorname{corr}(\Phi_X\mathbf{w}_X$, $\Phi_Y\mathbf{w}_Y)$, is
maximized. Note
that because $\mathbf{w}_X$ and $\mathbf{w}_Y$ lie in the span of
$\Phi_X$ and~$\Phi_Y$, these can be re-expressed by the linear transformations
$\mathbf{w}_X = \Phi_X\alpha_X$ and $\mathbf{w}_Y = \Phi_Y\alpha
_Y$. Letting
$\mathbf{K}_X = \Phi_X\Phi_X^T$ and $\mathbf{K}_Y = \Phi_Y\Phi
_Y^T$ with $k_X$
and $k_Y$ being the associated kernel functions for each space, respectively,
the CCA optimization problem in (\ref{EQCCA}) now becomes
%
\begin{eqnarray}\label{EQKCCA}
&&  \hspace*{35pt}\rho_{\mathcal{H}}  =
\max_{\mathbf{w}_X,\mathbf{w}_Y}\operatorname{corr}
(\Phi_{X}\mathbf{w}_X,\Phi_{Y}\mathbf{w}_Y) =
\max_{\alpha_X,\alpha_Y}\operatorname{corr}
(\mathbf{K}_X\alpha_X,\mathbf{K}_Y\alpha_Y)\nonumber
\\
&&  \hspace*{35pt}\mbox{subject to}\\
&&  \hspace*{35pt}\mathbf{w}_X^T\Phi_X^T\Phi_X\mathbf{w}_X = \alpha_X^T\mathbf
{K}_X^2\alpha_X =
1  \ \ \mbox{and} \ \ \mathbf{w}_Y^T\Phi_Y^T\Phi_Y\mathbf{w}_Y =
\alpha_Y^T\mathbf{K}_Y^2\alpha_Y = 1.\nonumber
\end{eqnarray}
 Here the subscript $\mathcal{H}$ in $\rho_{\mathcal{H}}$
is included
to emphasize the fact that the space of functions we are considering
are in a
RKHS. Subsequent directions are found by including the additional constraints
that $\alpha_X^{iT}\mathbf{K}_X^2\alpha_X^j =
\alpha_Y^{iT}\mathbf{K}_Y^2\alpha_Y^j =
\alpha_X^{iT}\mathbf{K}_X\mathbf{K}_Y\alpha_Y^j = 0$ for $i\neq j$, and
$\alpha_X^{iT}\mathbf{K}_X^2\alpha_X^i = \alpha_Y^{iT}\mathbf
{K}_Y^2\alpha_Y^i
= 1$, $i,j=1,\ldots,n$.

In order to avoid trivial solutions, we penalize the directions $\alpha
_X$ and
$\alpha_Y$ modifying the constraints in (\ref{EQKCCA}) to be
%
\begin{equation}\label{EQRKCCA}
\alpha_X^T\mathbf{K}_X^2\alpha_X + \kappa\alpha_X^T\alpha_X =
\alpha_Y^T\mathbf{K}_Y^2\alpha_Y + \kappa\alpha_Y^T\alpha_Y = 1.
\end{equation}
 Here $\kappa$ is a regularization parameter.

Note that the geometric interpretation of (unregularized) KCCA,
provided that
data have been centered in feature space, is the same as CCA. The only
difference lies in the fact that the space in which this geometry is observed
is in feature space rather than object space.

In order for KCCA to be understood as maximizing correlation in feature
space centering must be performed in feature space. Centering in
feature space
can be done as follows. Let $\bar{\Phi} = \frac{1}{n}\mathbf{J}\Phi
$ where
$\mathbf{J}$ is an $n\times n$ matrix of ones, then
\begin{eqnarray*}
(\Phi- \bar{\Phi})(\Phi- \bar{\Phi})^T
& =& \Phi\Phi^T - \Phi\bar{\Phi}^T - \bar{\Phi}\Phi^T + \bar
{\Phi}\bar{\Phi}^T \nonumber\\
& =& \mathbf{K} - \frac{1}{n}\mathbf{J}\mathbf{K} -
\frac{1}{n}\mathbf{K}\mathbf{J} + \frac{1}{n^2}\mathbf{J}\mathbf
{K}\mathbf{J}
\nonumber\\ & =& \biggl (\mathbf{I} - \frac{1}{n}\mathbf{J} \biggr)
\mathbf{K}
\biggl (\mathbf{I} - \frac{1}{n}\mathbf{J} \biggr).
\end{eqnarray*}
 We assume throughout that the kernel matrices are centered.

\begin{figure}

\includegraphics{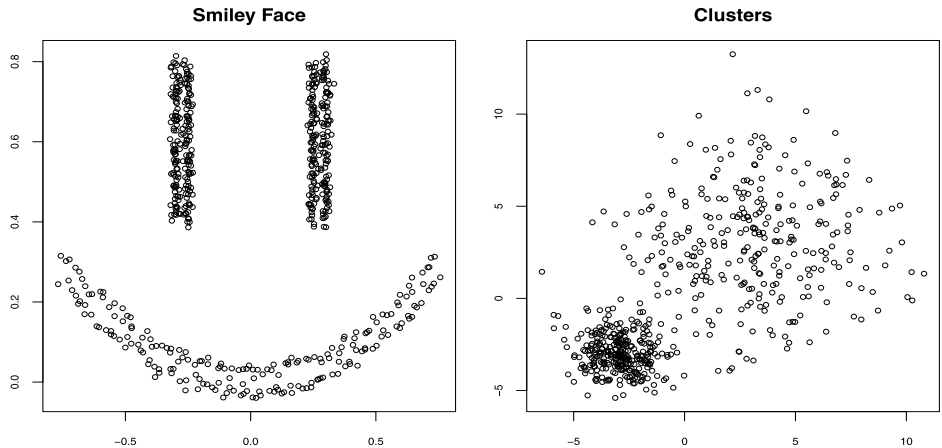}

\caption{A toy example illustrating the cases when the distribution of
points within a space is nonstandard and heterogeneous.}\vspace*{-1pt}
\label{smileydata}
\end{figure}

The predictive accuracy of this approach was discussed in Section
\ref{SECINTRODUCTION}, with results summarized in Figure \ref
{FIGIKCCA}. Recall
that the cyan line in Figure \ref{FIGIKCCA} corresponds to the
average predicted
rank using KCCA which is an improvement over both \citet{colibri2006}
and CCA.

\subsection{Toy example 3: KCCA challenge}
\label{SECTOYEXNONSTD}
We saw in Section \ref{exfeatmap} that KCCA was able to overcome
some of the
obstacles encountered by standard CCA. Where KCCA begins to encounter problems
is when the distribution of points within a space is nonstandard and/or
heterogeneous. To illustrate this consider the example shown in Figure
\ref{smileydata}, as with the protein--ligand matching problem, there
is a
one-to-one correspondence between points in the two spaces.\vadjust{\goodbreak}

\begin{figure}

\includegraphics{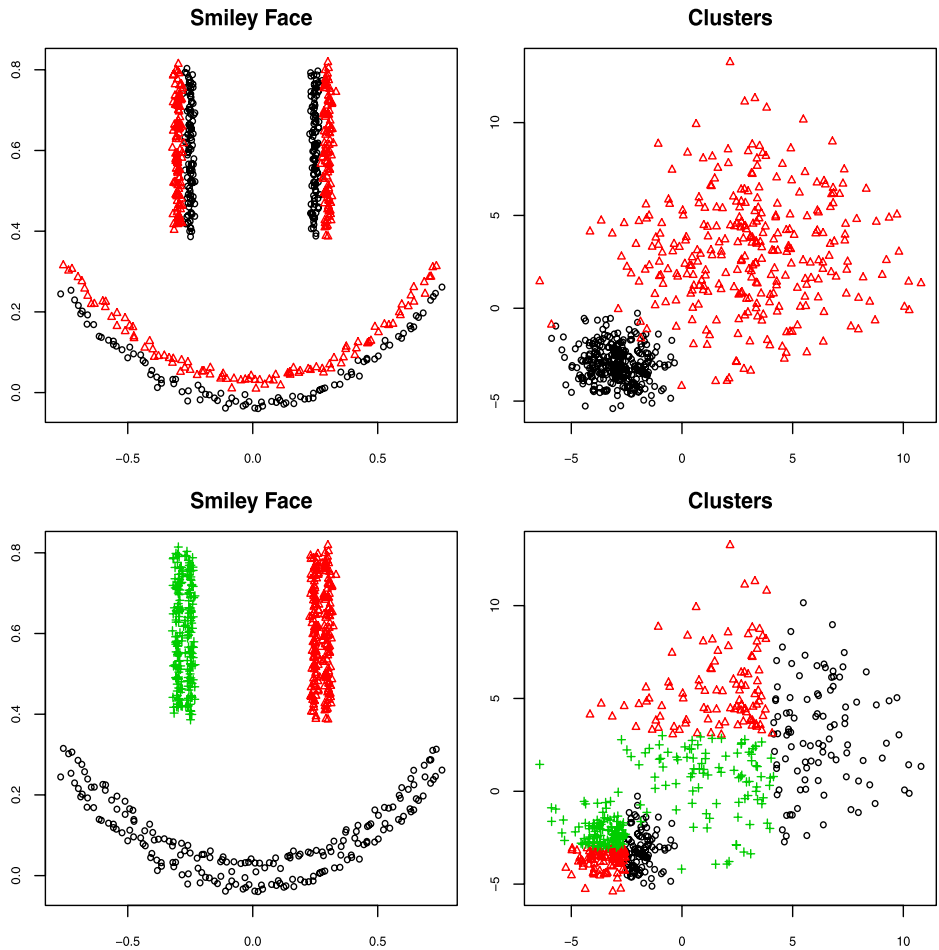}

\caption{These plots highlight how the distribution of points in one
space is related to the distribution of points in the other. Looking at the
plots on the left in Figure \protect\ref{smileystructure}, each of the three clusters
is in fact composed of two subclusters. Likewise, each of the two
clusters in
the plots on the right are composed of three subclusters.}
\label{smileystructure}
\end{figure}

\begin{figure}

\includegraphics{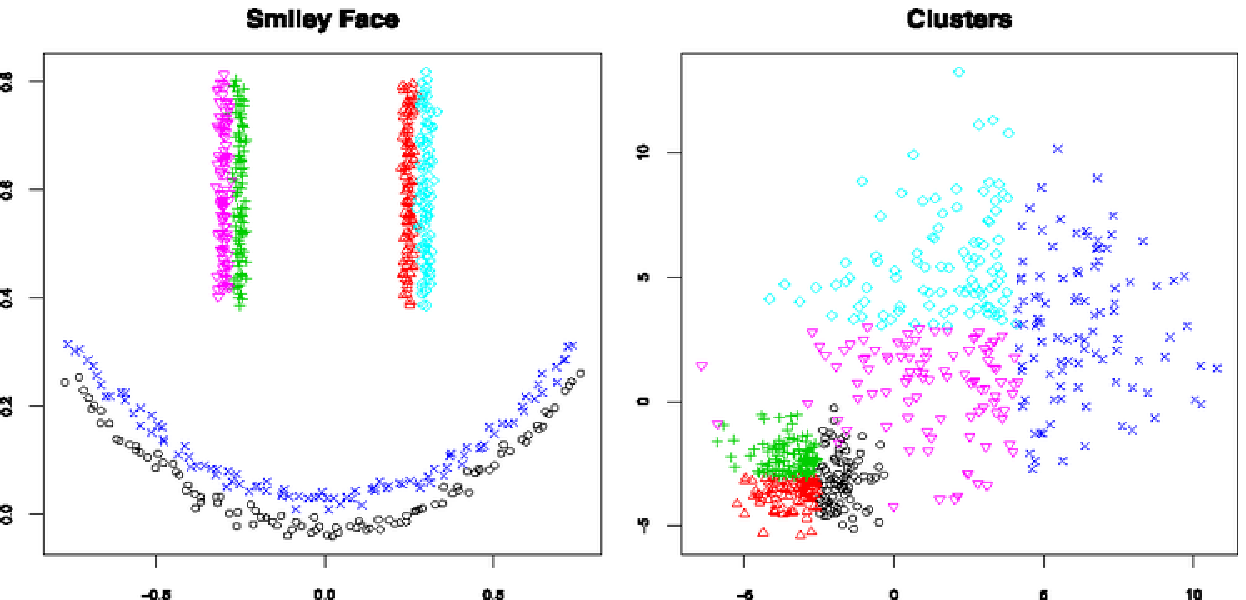}

\caption{In this plot each of the six underlying subgroups shown in
Figure \protect\ref{smileystructure} is highlighted.}
\label{FIGSMILEY6COLS}
\vspace*{5pt}
\end{figure}

The underlying structure between these spaces is illustrated in Figure
\ref{smileystructure}. The top row of plots tells us about how the
distribution of points on the right (cluster space) relates to the
distribution of points on the left (smiley face space). The bottom set of
plots tells us about how the distribution of points on the left is
related to
distribution of points on the right.

If we were to look at the two spaces as marginal distributions, there
is a
distinct impression of the three clusters in the left, and two in the right.
The joint distribution, however, has six distinct groups. Looking at
the plots
on the left in Figure \ref{smileystructure}, each of the three
clusters is in
fact composed of two subclusters. Likewise, each of the two clusters in the
plots on the right are composed of three subclusters. Ideally, the projections
onto the KCCA directions would identify each of these six groups, shown in
Figure \ref{FIGSMILEY6COLS}.

Using an RBF kernel with $\sigma= 1/2$ we look at the first five canonical
directions. Ideally, what we would see is a separation of each of the
groups as
well as a strong alignment between each of the spaces. What we find
looking at Figure \ref{smileyrbf}, a~scatter plot matrix of the first five kernel
canonical variates (KCV), is that while the leading correlations are large
(0.98, 0.97, 0.95, 0.80, 0.75), we are not able to find the structure
in the data we were looking for, that is, separating out the six groups (with
each of the colors corresponding to one of the six groups). Note that only the projections
in the smiley face space are shown since the cluster space projections look
essentially the same.

\begin{figure}

\includegraphics{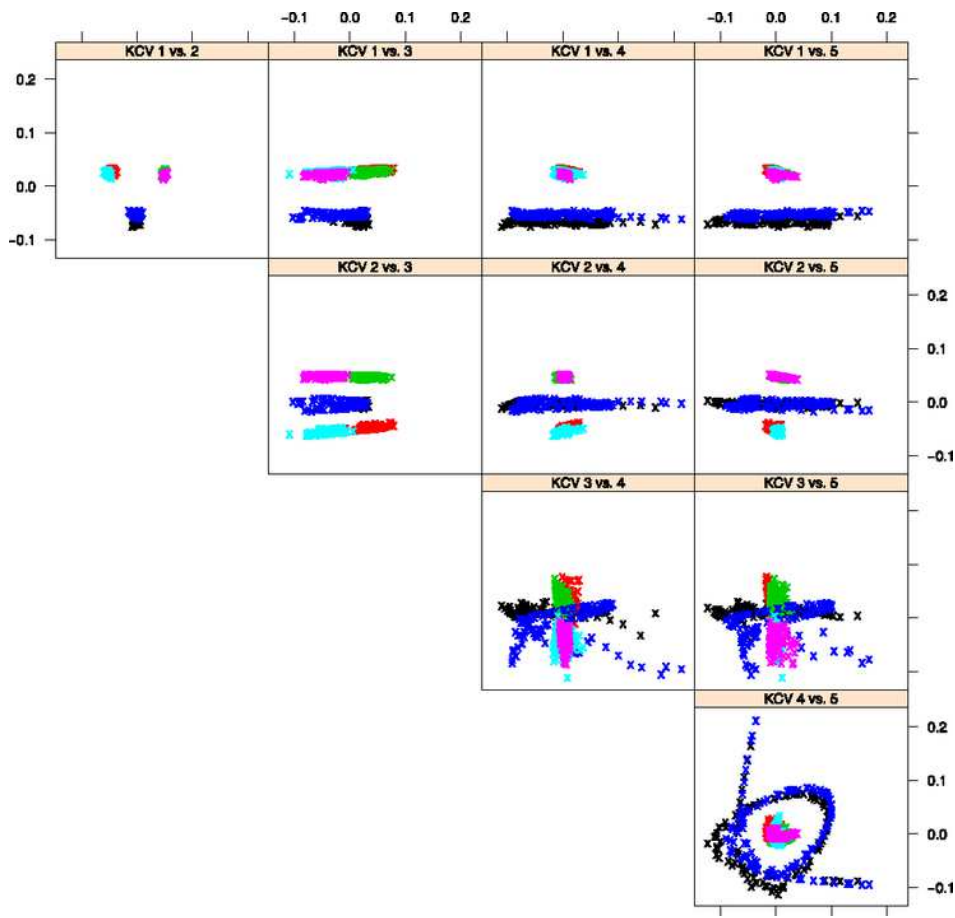}

\caption{Scatterplot matrix of the first five KCCA direction vectors for
the data shown in Figure~\protect\ref{smileydata}. Each of the colors in this plot
corresponds to one of the six underlying subpopulation in the data (see
Figure \protect\ref{smileystructure} for details).}
\label{smileyrbf}
\end{figure}

In the context of the protein--ligand matching problem this type of situation
presents a potential problem. Suppose a new point, say in the space
with the
smiley face, is projected into KCCA space. As can be seen in Figure
\ref{smileyrbf}, there is a great deal of overlap between each of the six
subgroups in the projected space. In particular note that each of the
overlapped groups is composed of, respectively, the left eye, right eye and
mouth. The reason this type of behavior presents a problem is that each
of the
eyes and the mouth are actually composed of two different
subpopulations where
each of the populations correspond to very different groups in the
space with
the two clusters. So while we may be able to accurately predict the location
of a new point in KCCA space the interpretation of its surrounding neighbors
may not be so meaningful.

\section{Indefinite kernel canonical correlation analysis}
\label{SECIKCCA}
A potential shortcoming of standard KCCA, which was illustrated in the example
presented in Figure~\ref{smileydata}, is that standard positive definite
kernels can be limited in their ability to capture nonstandard heterogeneous
behavior in the data. A more general class of kernels which is better
suited to
handle this type of behavior takes the form
%
\begin{equation}\label{localkernel}
K(\mathbf{x}_i,\mathbf{x}_j) =  \cases{\displaystyle
w(\mathbf{x}_i,\mathbf{x}_j), & \quad if   $\mathbf{x}_j\in
N(\mathbf{x}_i)$,\cr
0, & \quad otherwise.
}
\end{equation}
 Here $N(\mathbf{x})$ denotes some neighborhood of the observation
$\mathbf{x}$, such as a $k(\in\mathbb{Z}_+)$ nearest neighborhood or
a fixed
radius $\epsilon( > 0)$-neighborhood. Kernels of this form restrict
attention to
the local structure of the data and allow for a flexible definition of
similarity.

Our motivation for considering this class of kernels in the context of
the protein--ligand matching problem is the following. In the RLP800 dataset
there are approximately 150 important subgroups in the data. These subgroups
correspond to unique proteins, or more specifically their binding pockets,
which typically have three or four different conformations
specific to a particular ligand.
Exploitation of this group structure in the data can help improve prediction.
This can be accomplished by
using a ``local kernel'' function that allows us to capture these
groups more readily than, say, the RBF kernel. The intuition here
follows from
the example presented in Section \ref{SECTOYEXNONSTD} where we saw
that the
type of groups that an RBF kernel will be able to find will be dictated
by the
choice of the bandwidth parameter $\sigma$. The local kernel overcomes
this by
adjusting locally to the data. By adjusting to the data locally it is better
able to exploit this group structure.\looseness=1

In summary, given a new protein, its
projection in this local kernel CCA space will be more likely to fall
into a
group of similar proteins. Then, as before, the goal is that the ligands
associated with this group of proteins provide an accurate
representation of
the ligand we are trying to predict.

This improved performance exploiting group structure in the data comes
at some
price. In particular, the problem encountered with this class of
kernels is that
they are frequently indefinite (see the discussion following Definition
\ref{innprod}). As a result of the indefiniteness, many of the
properties and
optimality guarantees no longer hold.

Indefinite kernels have recently gained increased interest
[\citet{Ong04Krein}, \citet{Haasdonk2005SVM}, \citet{Chen2008SVM},
\citet{Luss2008SVM}] where, rather than defining $K$ to be a function defined
in a RKHS, $K$ is defined in a space characterized by an {\textit{indefinite
inner product}} called a~{\textit{Krein}} space. In Section~\ref{SECINDEFKERN}
we provide an overview of some of the definitions and theoretical results
about Krein spaces [following the discussion of \citet{Ong04Krein}].

Before discussing IKCCA, we will need to provide some definitions and theorems
related to indefinite inner product spaces, that is, Krein spaces [more details
can be found in \citet{Ong04Krein}].

\subsection{Indefinite kernels}
\label{SECINDEFKERN}
%
\begin{defn}[(Inner product)]
\label{innprod}
Let $\mathcal{K}$ be a vector space on the scalar field. An inner product
$\langle\cdot, \cdot \rangle_{\mathcal{K}}$ on $\mathcal{K}$ is a bilinear
form where
for all $f,g,h\in\mathcal{K}$, $\alpha\in\mathbb{R}$:
\begin{itemize}
\item$\langle f, g\rangle_{\mathcal{K}} = \langle g, f\rangle
_{\mathcal{K}}$;
\item$\langle\alpha f + g, h \rangle_{\mathcal{K}} = \alpha\langle
f, h
\rangle_{\mathcal{K}} + \langle g, h \rangle_{\mathcal{K}}$;
\item$\langle f, g\rangle_{\mathcal{K}} = 0$ for all $g\in\mathcal{K}$
implies $\Rightarrow f=0$.
\end{itemize}
\end{defn}
 The importance of $\mathcal{K}$ being a vector space on a
\textit{scalar field} is that it allows for a flexible definition of an inner
product (i.e., the scalar in one of the dimensions could be complex or negative
as we will see below). An inner product is said to be \textit{positive}
if for
all $f\in\mathcal{K}$, $\langle f, f \rangle_{\mathcal{K}} \geq0$.
It is
called a~\textit{negative} inner product, if for all $f\in\mathcal
{K}$, $\langle
f, f \rangle_{\mathcal{K}} \leq0$. An inner product is called
indefinite if
it is neither strictly positive nor strictly negative.

\begin{rem}
\label{REM1}
To illustrate how indefinite inner products arise in the
context of our problem, consider the following. Suppose we have a symmetric
kernel function $K$, which is indefinite, the implication of this is
that the
resulting kernel matrix $\mathbf{K} = \{K_{ij}\}_{i=1}^n$ is
indefinite and
that it therefore contains positive \textit{and} negative eigenvalues. Let
$\mathbf{K} = \mathbf{U}\bolds{\Lambda}\mathbf{U}^T$ be the
eigendecomposition of $\mathbf{K}$, where $\mathbf{U}$ are the eigenvectors
and $\bolds{\Lambda}$ is the diagonal matrix of eigenvalues starting
with the
$p$ positive eigenvalues, followed by the $q$ negative ones and the $n-p-q$
eigenvalues equal to 0. To see how $\mathbf{K}$ can be interpreted as
a matrix
composed of inner products in this indefinite inner product space
consider the
following representation of its eigendecomposition:
\[
\mathbf{K} =
\mathbf{U}|\bolds{\Lambda}|^{\fraca{1}{2}}\operatorname{diag}(\mathbf{1}_p,
-\mathbf{1}_q, \mathbf{0}_{n-p-q})|\bolds{\Lambda}|^{\fraca
{1}{2}}\mathbf{U}^T.
\]
 Let $\mathbf{M} = \operatorname{diag}(\mathbf{1}_p, -\mathbf
{1}_q)$ and
$\bolds{\Phi}$ be equal to the first $p + q$ columns of
$\mathbf{U}|\bolds{\Lambda}|^{\fraca{1}{2}}$. Define the $i${th}
row of
$\bolds{\Phi}$ to be equal to
\[
\Phi_i = (\underbrace{\phi_{i,1},\ldots, \phi_{i,p}}_{=\Phi_i^{+}},
\underbrace{\phi_{i,p+1}, \ldots, \phi_{i,p+q}}_{=\Phi_i^{-}}).
\]
We then have a kernel matrix composed of elements
%
\begin{eqnarray}
\label{EQIK}
K_{ij}
& =& \Phi_i^T\mathbf{M}\Phi_j \nonumber\\
& =& (\Phi_i^{+})^T\Phi_j^{+} - (\Phi_i^{-})^T\Phi_j^{-} \nonumber
\\[-8pt]
\\[-8pt]
& =& \langle\Phi_i, \Phi_j \rangle_{\mathcal{H}_{+}} - \langle\Phi
_i, \Phi_j
\rangle_{\mathcal{H}_{-}} \nonumber\\
& =& \langle\Phi_i, \Phi_j \rangle_{\mathcal{K}}.
\nonumber
\end{eqnarray}
 From (\ref{EQIK}) we can see that unlike PSD kernels where
$\mathbf{a}^T\mathbf{K}\mathbf{a}\geq0$ for any
\mbox{$\mathbf{a}\in\mathbb{R}^n$},
with indefinite kernels $\mathbf{a}^T\mathbf{K}\mathbf{a}$ can take
on any
value, making optimization over such a quantity challenging.

Despite this difference, many of the properties that hold for
reproducing kernel Hilbert spaces (RKHS), such as (and perhaps most importantly)
the reproducing property [\citet{learnWithKern2002}], also hold for these
indefinite inner product spaces [see \citet{ong2004} for details].
The key difference lies in the fact that rather than minimizing (maximizing)
a regularized risk functional, as in the RKHS setting, the corresponding
optimization problem becomes that of finding a stationary point of a similar
risk functional.
\end{rem}

%
\subsection{Indefinite kernel canonical correlation}
\label{SUBSECIKCCA}
%
Section \ref{SECINDEFKERN} provided some insight into the
challenges that arise from dealing with indefinite kernels. In particular,
Remark \ref{REM1} points to the fact that the solution that
we find may not be globally, or even locally, optimal (as it may be a saddle
point). The form of the IKCCA problem we present in this
section is motivated by the discussion of the previous section and the
works of
\citet{ong2004} and \citet{Luss2008SVM}.
In particular, the addition of a~stabilizing function on the indefinite inner product $\langle f, f
\rangle_{\mathcal{K}}$ as discussed in \citet{ong2004} led us to consider
introducing a constraint on the behavior on the indefinite kernels
matrix itself.

In the following, let $\|\cdot\|_F$ denote the Frobenius norm. Define
$\mathbf{M}\succeq0$ to mean that the matrix $\mathbf{M}$ is positive
semi-definite and let $\lambda_X, \lambda_Y\in\mathbb{R}^{+}\cup
\infty$ be
tuning parameters (discussed in more detail later this section). Here
$\mathbf{K}_X^0$ and $\mathbf{K}_Y^0$ are the (potentially)
indefinite kernels
and $\mathbf{K}_X$ and $\mathbf{K}_Y$ will be the positive semi-definite
approximations of these kernels. With this notation in mind, we now define
the IKCCA optimization problem:
%
\begin{eqnarray}
\label{sccaoptim}
&& \rho_{\mathcal{H}} = \max_{\mathbf{A}_X,
\mathbf{A}_Y}\min_{\mathbf{K}_X,\mathbf{K}_Y}
\operatorname{Tr}(\mathbf{A}_X^T\mathbf{K}_X \mathbf{K}_Y\mathbf{A}_Y) +
\lambda_X\|\mathbf{K}_X - \mathbf{K}_X^0\|_F^2\nonumber\\
&&\hphantom{\rho_{\mathcal{H}} =}{} + \lambda_Y\|\mathbf
{K}_Y -
\mathbf{K}_Y^0\|_F^2, \nonumber\\
&&\mbox{subject to}\nonumber\\
&&\mathbf{A}_X^T\mathbf{K}_X^2\mathbf{A}_X + \kappa\mathbf
{A}_X^T\mathbf{A}_X =
\mathbf{I}_n, \nonumber\\
&& \mathbf{A}_Y^T\mathbf{K}_Y^2\mathbf
{A}_Y +
\kappa\mathbf{A}_Y^T\mathbf{A}_Y = \mathbf{I}_n,
\\
&&
(\alpha_X^i)^T\mathbf{K}_X\mathbf{K}_Y\alpha_Y^j = 0 \qquad \mbox{for }
i\neq j,i,j=1,\ldots,n, \nonumber\\
&& \mathbf{K}_X \succeq\mathbf{0},
\nonumber\\
&& \mathbf{K}_Y \succeq\mathbf{0},
\nonumber
\end{eqnarray}
 where $\mathbf{A}_X = (\alpha_X^1,\ldots,\alpha_X^n)$ and
$\mathbf{A}_Y = (\alpha_Y^1,\ldots,\alpha_Y^n)$. Note that this optimization
problem and the KCCA optimization problem are only equivalent when the kernel
matrices $\mathbf{K}_X^0$ and $\mathbf{K}_Y^0$ are positive semi-definite
(see the \hyperref[samarovTech2010]{Supplementary Material} [\citet{Sam2011}] for details on the
equivalency between the
optimization problem in (\ref{sccaoptim}) and (\ref{EQKCCA}) and a
proof of
Theorem \ref{THMIKCCAKOPTIM}).

\begin{theo}
\label{THMIKCCACAVVEX}
Letting $\lambda_X,\lambda_Y\rightarrow\infty$, the optimization
problem in
$(\ref{sccaoptim})$ is concave in $\alpha_X^i$ and $\alpha_Y^i$, $i =
1,\ldots,n,$ and convex in $\mathbf{K}_X$ and $\mathbf{K}_Y$.
\end{theo}

 See the \hyperref[samarovTech2010]{Supplementary Material} for a proof.
Let $(\mathbf{X})_{+}$ denote the positive part of the matrix $\mathbf
{X}$, that is, $(\mathbf{X})_+ =
\sum_i \max(0,\lambda_i)\mathbf{v}_i\mathbf{v}_i^T$, where
$\lambda_i$ and
$\mathbf{v}_i$ are $i$th eigenvalue--eigenvector pair of the matrix
$\mathbf{X}$. With this in mind, we have the following theorem.
%
\begin{theo}
\label{THMIKCCAKOPTIM}
Letting $\lambda_X,\lambda_Y\rightarrow\infty$, and given the optimization
problem in $(\ref{sccaoptim})$ the optimal values for $\mathbf{K}_X$ and
$\mathbf{K}_Y$ are given by
%
\begin{eqnarray}
\label{kkernel}
 \mathbf{K}_X &=& (\mathbf{K}_X^0)_{+}, \nonumber
\\[-8pt]
\\[-8pt]
 \mathbf{K}_Y &=& (\mathbf{K}_Y^0)_{+}.
\nonumber
\end{eqnarray}
\end{theo}

 The proof of Theorem \ref{THMIKCCAKOPTIM} makes use of the
following lemma. Let $\mathbf{M}_0\in\mathbb{R}^{n\times n}$ be a known,
square, not necessarily positive-definite matrix, and
$\mathbf{M}\in\mathbb{R}^{n\times n}$ a square, unknown matrix,
then:\vspace*{-3pt}
%
\begin{lem}
\label{LEMFROBOPTIM}
The solution to the optimization problem
\[
\label{EQLEMFROBOPTIM}
\operatorname{arg}\min_{\mathbf{M} \succeq0}\|\mathbf{M} -
\mathbf{M}_0\|^2_F,
\]
 is
\[
\mathbf{M} =  (\mathbf{M}_0 )_{+}.
\]
\end{lem}
%

 The proofs of Theorem \ref{THMIKCCAKOPTIM} and Lemma
\ref{LEMFROBOPTIM} can be found in the \hyperref[refid=samarovTech2010]{Supplemen-}
\hyperref[samarovTech2010]{tary Material}.

Points $\mathbf{x}\in\mathbb{R}^{d_X}$ and $\mathbf{y}\in\mathbb
{R}^{d_Y}$ are
projected onto their first $p$ canonical directions as follows: first compute
their kernelization, using the indefinite kernel functions $K_X^0$ and $K_Y^0$,
\begin{eqnarray*}
  K_X^0(\mathbf{x},\cdot) &=& (K_X^0(\mathbf{x},\mathbf{x}_1), \ldots,
K_X^0(\mathbf{x},\mathbf{x}_n))^T,\\
 K_Y^0(\mathbf{x},\cdot) &=&
(K_Y^0(\mathbf{y},\mathbf{y}_1), \ldots, K_Y^0(\mathbf{y},\mathbf{y}_n))^T.
\end{eqnarray*}
 Then calculate
\begin{eqnarray*}
  &&K_X^0(\mathbf{x},\cdot)\mathbf{A}_X^p, \\
  &&K_Y^0(\mathbf{y},\cdot)\mathbf{A}_Y^p,
\end{eqnarray*}
 where $\mathbf{A}_X^p = (\alpha_X^1,\ldots,\alpha_X^p)$ and
$\mathbf{A}_Y^p = (\alpha_Y^1,\ldots,\alpha_Y^p)$.

\vspace*{-4pt}\subsection{Toy example 3: KCCA challenge (motivating IKCCA)}
\label{SUBSECTOYEX}
We now return to the example in Section \ref{SECTOYEXNONSTD} using the
kernel defined in (\ref{localkernel}) with weights
(\ref{EQNGLWEIGHTSRBF}). Note that this kernel is closely related
to the
Normalized Graph Laplacian (NGL) kernel used in Spectral Clustering; see
\citet{vonLux2007Spec} for an overview of Spectral Clustering methods. From
Figure \ref{glsmiley}, it can be seen that we are now able to capture the
underlying structure of the data, identifying each of the six subpopulations:
\[
a_{ij} =
\cases{\displaystyle
\exp \biggl\{-\frac{1}{2\sigma_{ij}}\|\mathbf{x}_i - \mathbf
{x}_j\|^2 \biggr\}, & \quad if    $\mathbf{x}_j\in N_k(\mathbf
{x}_i)$, \cr\displaystyle
0, & \quad otherwise,
}
\]
 and
%
\begin{equation}
\label{EQNGLWEIGHTSRBF}
w_{ij} = \frac{a_{ij}}
{
\sqrt{\sum_{i^{\prime}}a_{i^{\prime}j}}
\sqrt{\sum_{j^{\prime}}a_{ij^{\prime}}}
}.
\end{equation}
 Here $N_k(\mathbf{x}_i)$ is the symmetric $k$-neighborhood
of the
point $\mathbf{x}_i$ [i.e., if $\mathbf{x}_j\in N_k(\mathbf{x}_i)$ then
$\mathbf{x}_i\in N_k(\mathbf{x}_j)$] and
\[
\sigma_{ij}^2 = \|\mathbf{x}_i - \mathbf{x}_i^k\|\|\mathbf{x}_j -
\mathbf{x}_j^k\|,
\]
 where $\mathbf{x}_i^k$ is the $k$th neighbor of the point
$\mathbf{x}_i$.
%
\begin{figure}

\includegraphics{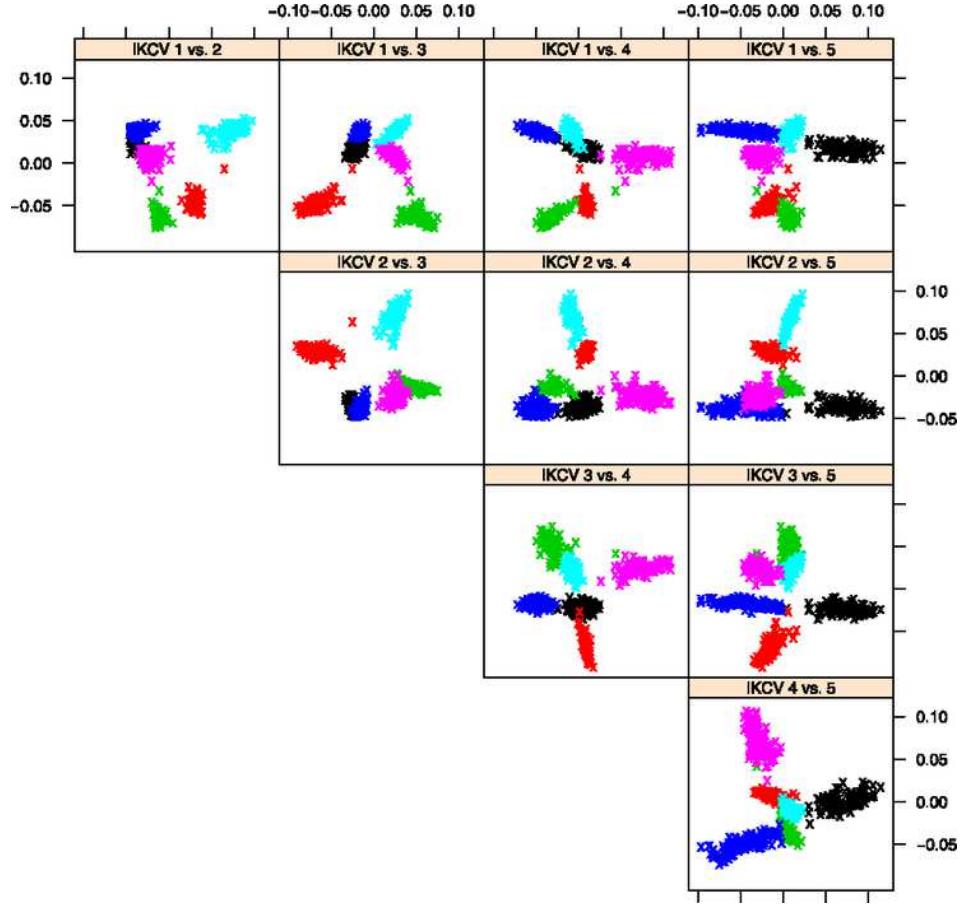}

\caption{Continuation from the example in Section \protect\ref{SUBSECTOYEX}.
This is a scatter plot matrix of the projections onto the first five IKCCA
variates (IKCV) using the kernel in (\protect\ref{localkernel}) with weights
(\protect\ref{EQNGLWEIGHTSRBF}). Unlike the projections shown in Figure
\protect\ref{smileyrbf}, here we are able to separate out the six groups.}
\label{glsmiley}
\end{figure}

\begin{figure}

\includegraphics{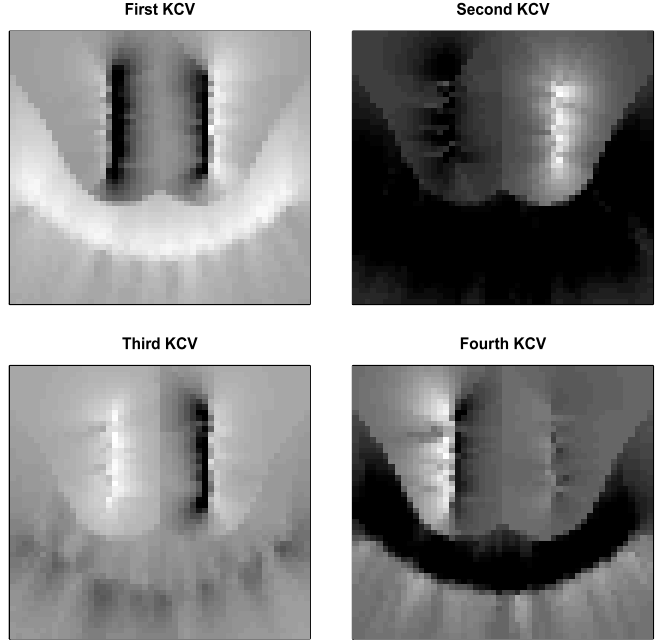}

\caption{A plot of the first four indefinite kernel canonical direction
vectors in the smiley face space from the example in Section
\protect\ref{SUBSECTOYEX} using the kernel in (\protect\ref{localkernel}) with weights
(\protect\ref{EQNGLWEIGHTSRBF}). These plots allow us to visualize how the
canonical vectors separate out each of the clusters.}
\label{eigensmile}
\vspace*{-5pt}
\end{figure}

\begin{figure}

\includegraphics{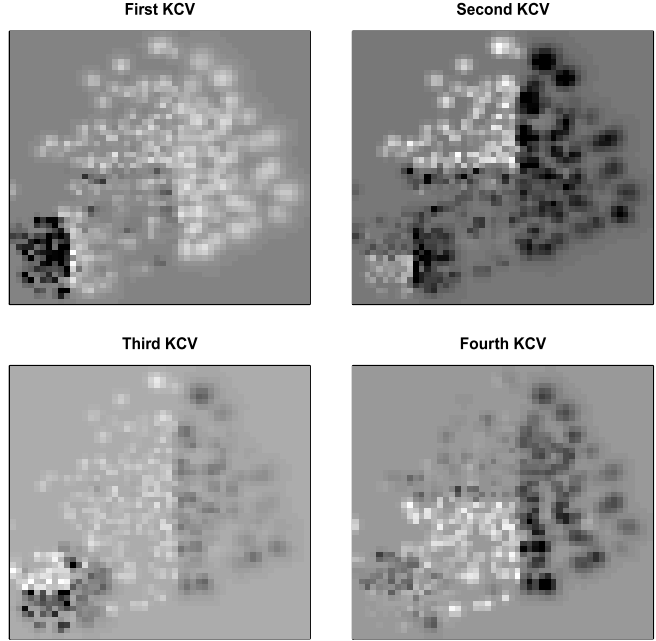}

\caption{A plot of the first four indefinite kernel canonical directions
vectors in the cluster space from the example in Section \protect\ref
{SUBSECTOYEX}
using the kernel in (\protect\ref{localkernel}) with weights
(\protect\ref{EQNGLWEIGHTSRBF}).}
\label{eigencluster}
\end{figure}

Looking at plots of the first four eigenvectors (Figures \ref
{eigensmile} and
\ref{eigencluster}) in both the smiley face space and the cluster
space, we
can see how the behavior of the eigenvectors causes the segmentation of the
data that we observe in Figure \ref{glsmiley}. First, we discuss how these
figures are generated and then what it is they are telling us.
\begin{longlist}[(3)]
\item[(1)] Generate an equally spaced dimensional grid spanning the range of values
in each space.
\item[(2)] Calculate the kernel representation and projection of each grid point
into IKCCA space.
\item[(3)] Use the projected values to assign color intensities to each
point in
the grid of each space (darker for negative values, lighter for positive
values).
\item[(4)] Plot the grid and for each point using the colors calculated from the
previous step.
\end{longlist}
 The important thing to note in both of these figures is the
distribution of positive and negative projected values and how these are
driving the segmentation, which we observe in Figure \ref{glsmiley}. For
example, in Figure \ref{eigensmile} the first canonical variate
segments out
one of the faces (red) from the other (blue).

\vspace*{-2pt}
\section{Ligand prediction}
\vspace*{-2pt}
\label{SECPREDICTION}

\subsection{Prediction}
\label{SUBSECPREDICTION}
Let us define the projected values of the observations in protein and ligand
space onto their first $p$ canonical vectors\vspace*{-1pt} as $\mathbf{x}_{i,p}^w =
 (\mathbf{w}_X^1,\ldots, \mathbf{w}_X^{p_X} )^T\mathbf{x}_i$,
$\mathbf{x}_{i,p}^w\in\mathbb{R}^p$, and $\mathbf{y}_{i,p}^w =
 (\mathbf{w}_Y^1,\ldots,\mathbf{w}_Y^{p_Y} )^T\mathbf{y}_i$,
$\mathbf{y}_{i,p}^w\in\mathbb{R}^p$,   $i = 1,\ldots,n$. The predicted
value of $\mathbf{y}^w_{\mathrm{new},p}$ is calculated as follows [using a modification
of the LLE algorithm of \citet{Saul03thinkglobally}]:
\begin{longlist}[(3)]
\item[(1)] Compute the $k$ neighbors of the data point $\mathbf
{x}^w_{\mathrm{new},p}$ (the
projected value of $\mathbf{x}_{\mathrm{new}}$ into canonical correlation space).
Define $N_k(\mathbf{x})$ to be the $k$ nearest neighbors of the point
$\mathbf{x}$.
\item[(2)] Compute weights $\beta_{\mathrm{new},j}$ that best reconstruct the data point
$\mathbf{x}^w_{\mathrm{new},p}$ from its neighbors, minimizing the cost
%
\begin{eqnarray}
  &&L(\beta_{\mathrm{new}}) =  \biggl(\mathbf{x}^w_{\mathrm{new},p} - \sum_{j\dvtx \mathbf
{x}_j\in
N_k(\mathbf{x}^w_{\mathrm{new},p})}\beta_{\mathrm{new},j}\mathbf{x}^w_{j,p}
\biggr)^2,\nonumber\\
&&\mbox{subject to}\\
 &&\sum_{j\dvtx \mathbf{x}_j\in
N_k(\mathbf{x}^w_{\mathrm{new},p})}\beta_{\mathrm{new},j} = 1.\nonumber
\end{eqnarray}
\item[(3)] The new observation is then calculated as
\[
\hat{\mathbf{y}}^w_{\mathrm{new},p} = \sum_{j\dvtx \mathbf{x}_j\in
N_k(\mathbf{x}^w_{\mathrm{new},p})}\beta_{\mathrm{new},j}\mathbf{y}^w_{j,p}.
\]
\end{longlist}
 Recall that CCA finds directions which best align two spaces. Thus,
assuming that directions $\mathbf{w}_X^i$ and $\mathbf{w}_Y^i$, $i =
1,\ldots,p$, have been found such that the correlation between spaces is
strong, using the weights $\beta_{\mathrm{new},j}$ found in protein space should
provide a reliable estimate of
$\mathbf{y}^w_{\mathrm{new},p}$.

\subsection{Tuning parameter selection}
\label{SUBSECTUNEPARAMS}
Values for the tuning parameters, $\kappa= \kappa_X = \kappa_Y$ (the
regularization parameter), $p = p_X = p_Y$ (the number of dimensions we are
projecting into), $k_{\mathrm{LLE}}$ (the neighborhood for the LLE-based prediction),
$\sigma$ (for the RBF kernel) and $k_{\mathrm{NGL}}$ (for the NGL kernel) are
found by
searching over a suitable $3\times3\times3\times3$ grid for each.
The final
set of parameters are selected based on which produces the lowest
average rank
(discussed in Section \ref{SECINTRODUCTION}).

\begin{supplement}[id=samarovTech2010]
\stitle{Local kernel canonical correlation analysis with application
to virtual
drug screening}
\slink[doi]{10.1214/11-AOAS472SUPP} 
\slink[url]{http://lib.stat.cmu.edu/aoas/472/supplement.pdf}
\sdatatype{.pdf}
\sdescription{Drug discovery is the process of identifying compounds
which have potentially
meaningful biological activity. A~major challenge that arises is that the
number of compounds to search over can be quite large, sometimes
numbering in
the millions, making experimental testing intractable. For this reason
computational methods are employed to filter out those compounds which
do not
exhibit strong biological activity. This filtering step, also called virtual
screening reduces the search space, allowing for the remaining
compounds to be
experimentally tested.\\
\indent
In this paper we propose several novel approaches to the problem of virtual
screening based on Canonical Correlation Analysis (CCA) and on a~kernel based
extension. Spectral learning ideas motivate our proposed new method called
Indefinite Kernel CCA (IKCCA). We show the strong performance of this approach
both for a toy problem as well as using real world data with dramatic
improvements in predictive accuracy of virtual screening over an existing
methodology.}
\end{supplement}

%

\printaddresses

\end{document}